\documentclass[runningheads]{llncs}

\usepackage[T1]{fontenc}
\usepackage{graphicx}

\begin{document}

\title{Congestion-Based Slot Pricing in a Railway Auction Game}

\author{Bill Roungas\inst{1,2} \and
Sebastiaan Meijer\inst{3}}

\authorrunning{B. Roungas et al.}

\institute{Panteion University, Athens, Greece \and
Institute of Communication and Computer Systems, Athens, Greece\\
\email{vroungas@panteion.gr} \and
KTH Royal Institute of Technology, Stockholm, Sweden\\
\email{smeijer@kth.se}}

\maketitle

\begin{abstract}
We present a multi-agent system for studying the allocation of discrete, congested resources among heterogeneous strategic agents, motivated by the problem of railway slot allocation under deregulation. Multiple operator-agents, differing in size and capacity, interact through a shared auction mechanism over repeated rounds under time-constrained decision-making. The mechanism combines a congestion-based base price that increases with aggregate demand with an asymmetric corrective adjustment that penalises the agent requesting the most slots and rewards the agent requesting the fewest, and is designed to mitigate strategic dominance by large agents while preserving transparency and congestion sensitivity. We formulate the interaction as a repeated game with incomplete information and implement the system as a real-time, web-based multi-agent environment in which human participants control individual agents and observe live marginal-cost and competitor feedback.

We report exploratory observations from two structured sessions with domain experts acting as operator-agents. The congestion mechanism responds to aggregate demand as designed and the corrective incentives are actively triggered, but agents representing large operators persist with high-request strategies despite the penalty, suggesting that corrective pricing is necessary but not sufficient to neutralise strategic dominance in this multi-agent setting. A post-session debrief indicates that participants' decisions were driven by the assumed agent role rather than personal disposition, and provides qualitative support for strategic motives, such as preserving market presence and raising rivals' costs, operating alongside short-term profit maximisation. We discuss implications for multi-agent mechanism design under asymmetric budgets and outline directions for analytical validation and larger-scale multi-agent experiments.

\end{abstract}

\section{Introduction}\label{sec:intro}

The allocation of scarce, congested resources among multiple strategic agents with heterogeneous capabilities and objectives is a recurring challenge in the design of multi-agent systems. Railway capacity allocation under deregulation provides a particularly clear instance of this problem. Over the past decades, railway systems in many Western countries have opened train operations to competition while retaining the underlying infrastructure in public ownership. This institutional separation creates a multi-agent setting in which train operators (who are agents that differ substantially in size, financial capacity and service portfolios) compete for discrete slots on a shared infrastructure managed by a public authority. Unlike private-market resource allocation, the authority is required to balance economic efficiency, competition and societal welfare rather than pursue profit maximisation alone \cite{ait2020pricing}, which places distinctive constraints on the mechanisms that can govern the interaction.

One of the most persistent challenges arising from deregulation is the management of conflicting capacity requests from multiple operators, particularly in high-demand corridors and during peak periods. Train paths, or slots, constitute discrete and scarce resources, and their allocation has significant implications for market structure, service quality, and long-term competition \cite{schlechte2010railway}. Traditional administrative allocation mechanisms often lack transparency and may fail to provide appropriate incentives for efficient use of capacity. As a result, market-based approaches, including auctions, have received increasing attention as potential tools for railway capacity allocation \cite{borndorfer2006auctioning,harrod2013auction}.

Pricing railway slots through auctions, however, raises concerns that extend beyond allocative efficiency. In markets where operators differ substantially in size and financial capacity, auction mechanisms may enable dominant firms to crowd out smaller competitors, either by consistently bidding for excessive capacity or by temporarily absorbing losses to raise prices. For publicly governed railway systems, such outcomes are undesirable, as they risk reducing competition and undermining societal welfare. Consequently, auction mechanisms for railway capacity must be carefully designed to discourage monopolistic practices while maintaining transparency and congestion sensitivity.

This paper extends a previously proposed auction-based model for pricing railway slots \cite{roungas2021auction,broman5007501track} by explicitly incorporating corrective incentives aimed at mitigating the strategic advantages of large operators. The model introduces congestion-based pricing, whereby the cost of slots increases with total demand, alongside a reward-and-penalty mechanism that favours operators requesting relatively fewer slots. To support the design and evaluation of the proposed auction model, a real-time, multiplayer game was developed, allowing participants to repeatedly request slots under dynamically adjusting prices and time-constrained decision-making.

The primary contribution of this paper is twofold. First, it presents the design and implementation of a multi-agent auction game that operationalizes the proposed pricing mechanism in an interactive, transparent and observable manner. Second, it reports initial observations from structured gameplay sessions involving operators of different sizes. These observations provide exploratory insights into how participants respond to congestion pricing, corrective incentives, and time pressure, and serve as a preliminary empirical grounding for the proposed approach. A more extensive analytical and experimental evaluation is reserved for future work.

\section{Background Work}\label{sec:background}

The allocation of railway capacity is a central problem in railway planning and operations, as it directly affects timetable feasibility, market access, and service reliability. A conflict-free timetable is a fundamental requirement for railway systems, and auctions for railway capacity are therefore commonly modeled as auctions of discrete goods, namely train slots \cite{schlechte2010railway}. Due to their discrete nature and scarcity—particularly during peak periods—railway slots have been the focus of numerous auction-based allocation approaches.

Several studies have examined auction mechanisms aimed at improving the efficiency of railway capacity allocation. In \cite{borndorfer2006auctioning}, an auctioning algorithm is proposed in which operators bid for bundles of slots rather than individual slots, reflecting the interdependencies between train paths in a timetable. This bundle-based perspective is partially adopted in the present work, in the sense that operators submit a single bid for a set of slots in each round rather than bidding sequentially for individual slots. From a financial and regulatory perspective, Harrod \cite{harrod2013auction} examined auctions as a mechanism for allocating railway capacity in North America, highlighting both their potential efficiency gains and their suitability for regulated infrastructure markets.

Beyond efficiency considerations, transparency and accountability are recurring themes in the literature on railway capacity auctions. Auctions have been argued to enhance transparency by making allocation and pricing rules explicit and observable to all participants, which can contribute to improved societal welfare \cite{harrod2013auction}. Transparency is particularly important in publicly governed railway systems, where infrastructure managers are accountable not only for economic outcomes but also for fairness and long-term market structure.

At the same time, several authors have noted the risk that auction-based mechanisms may enable dominant operators to exploit their financial strength, potentially leading to monopolistic outcomes. In particular, pricing mechanisms that apply uniform price increases to all operators may be strategically manipulated by large firms willing to temporarily absorb losses in order to raise rivals’ costs \cite{borndorfer2006auctioning,schlechte2010railway}. Dual pricing schemes and congestion-sensitive pricing models have therefore been proposed as alternatives that better reflect network usage while limiting anti-competitive behavior \cite{schlechte2010railway}.

While much of the existing work focuses on analytical models and optimization-based solutions, there is comparatively less emphasis on how human decision-makers interact with such mechanisms in practice. Gaming simulations have been proposed as a means to bridge this gap by enabling experimentation with complex systems while capturing behavioural and strategic aspects of decision-making \cite{roungas2021improving,roungas2019game}. From a game science perspective, games function as temporary social systems consisting of interconnected actors, rules, and resources, where participants co-construct meaning through interaction \cite{meijer2025positioning}. Such approaches allow designers to explore how participants respond to incentives, uncertainty, and repeated interaction in controlled environments.

The present study builds on this body of work by embedding an auction-based pricing mechanism within a multiplayer game. Rather than aiming to derive an optimal allocation algorithm, the focus is on exploring how different incentive structures influence operator behavior over repeated rounds. This approach complements analytical validation methods \cite{roungas2017framework,roungas2018framework} by providing behavioural observations that can inform future refinement of both the auction model and the game design.

\section{Design Methodology \& Implementation}\label{sec:methodology}

\subsection{Auction Model}\label{subsec:model}

The auction model underlying the game is intentionally simplified and draws on concepts from game theory, as the primary objective is not to replicate the full complexity of real-world railway operations, but to explore how different pricing and incentive structures influence operator behavior \cite{roungas2019game}. The game is round-based and can, in principle, continue indefinitely, corresponding to a repeated game with incomplete information \cite{aumann1995repeated}. Each game session represents a specific operational scenario, defined by parameters such as origin–destination pair, time window (e.g. peak or off-peak hours), train types, and the characteristics of the participating operators, including their available rolling stock.

In each round, the only decision available to players is the number of slots they request. Based on this decision and the total number of slots requested by all operators, the model computes the resulting costs, revenues, and profits for each participant. Revenues are calculated using assumptions regarding passenger demand, ticket prices, and market shares, which are held constant within a game session. Costs consist of three components: the cost of acquiring slots, the operational cost of running trains, and the cost of renting trains when an operator’s own rolling stock is insufficient to operate all requested slots.

The core contribution of the auction model lies in the determination of slot acquisition costs. In all model variations, the price per slot increases as the total number of requested slots increases, reflecting congestion on the railway network. This price increase applies uniformly to all slots acquired by an operator. Such congestion-based pricing is intended to internalize the cost of excessive capacity usage and discourage inefficient over-demanding of slots.

Two variations of the slot pricing mechanism are considered. In the first variation, all operators pay the same price per slot, regardless of how many slots they individually request. While this approach treats operators symmetrically, it carries the risk that large operators may strategically request excessive capacity in order to raise slot prices for all participants, potentially disadvantaging smaller competitors.

To address this concern, a second variation introduces asymmetric corrective incentives. After the base slot price has been determined through congestion-based pricing, the operator requesting the highest number of slots is subject to a penalty on slot acquisition costs, while the operator requesting the lowest number of slots receives a corresponding discount. In the implementation used for gameplay, this penalty and discount amount to 20\% of the slot acquisition cost and do not apply to other cost components such as operational or rolling stock costs. The purpose of this mechanism is to discourage aggressive overbidding by large operators and to incentivize participation by smaller operators, even when requesting a limited number of slots.

The game can theoretically continue for an unlimited number of rounds, and the criteria for ending a game session are determined by the designer. Two possible termination conditions have been considered. The first is a fixed number of rounds, which provides predictability but allows participants to anticipate the end of the game and potentially adjust their strategies accordingly. The second is a convergence-based criterion, where the game ends when no operator changes its requested number of slots for two consecutive rounds. While this approach limits end-game strategising, it results in unpredictable game duration and poses practical challenges for coordinating game sessions.

\subsection{Prototype}\label{subsec:prototype}

\begin{figure}
\includegraphics[width=\textwidth]{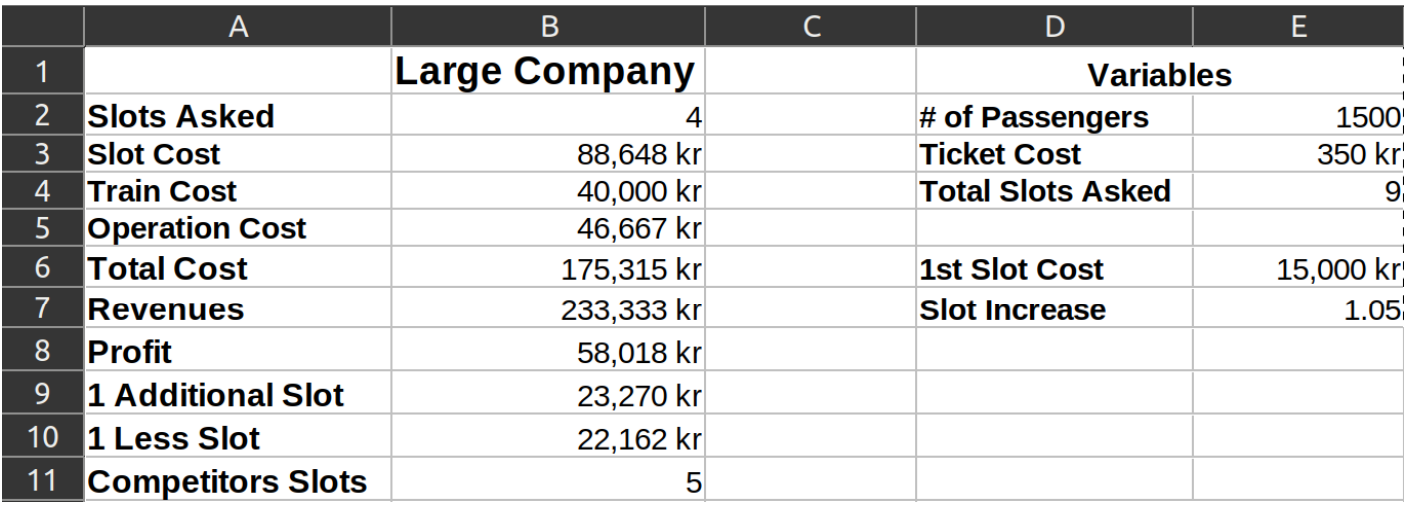}
\caption{The prototype.} \label{fig:prototype}
\end{figure}

To support the development of the auction model and to identify potential usability and design issues, an initial prototype of the game was implemented in Microsoft Excel (Figure \ref{fig:prototype}). Although limited in interactivity, the prototype was fully playable and served as an effective tool for testing the internal logic of the auction model and the financial calculations underlying the game.

In the prototype, players could modify only a single input: the number of slots requested in a given round. All other values, including costs, revenues, profits, and marginal cost indicators, were calculated automatically based on this input and the total number of slots requested by competitors. The prototype therefore allowed players to immediately observe how changes in their own slot requests would affect their financial outcomes, assuming competitors’ decisions remained unchanged.

The structure of the spreadsheet was designed to clearly separate inputs, calculated variables, and constants. With the exception of the total number of requested slots, which varied dynamically based on player input, most parameters were held constant within a game session. The total number of slots requested by competing operators was imported from separate spreadsheet files, allowing the prototype to simulate multiplayer interaction in a simplified manner.

Developing the prototype in Excel proved useful for validating the internal consistency of the auction model and for identifying features that would be impractical or unintuitive to implement in a spreadsheet environment. In particular, aspects such as real-time interaction, automatic synchronization between players, and time-constrained decision-making highlighted the need for a more robust implementation. These insights informed the design of the final, web-based version of the game, which is described in the following section.

\subsection{Implementation}\label{subsec:implementation}

The final version of the game was implemented as a web-based application using standard web technologies, including HTML, CSS, JavaScript, PHP, and a MySQL database. This implementation enabled real-time multiplayer interaction, automated data collection, and more flexible user interface design compared to the spreadsheet prototype.

After logging into the system, participants are guided through a sequence of instruction pages that explain the rules of the game, the auction mechanism, and the financial logic underlying the displayed values. At the end of this introduction phase, players submit their initial slot request, after which the game proceeds in rounds.

\begin{figure}
\includegraphics[width=\textwidth]{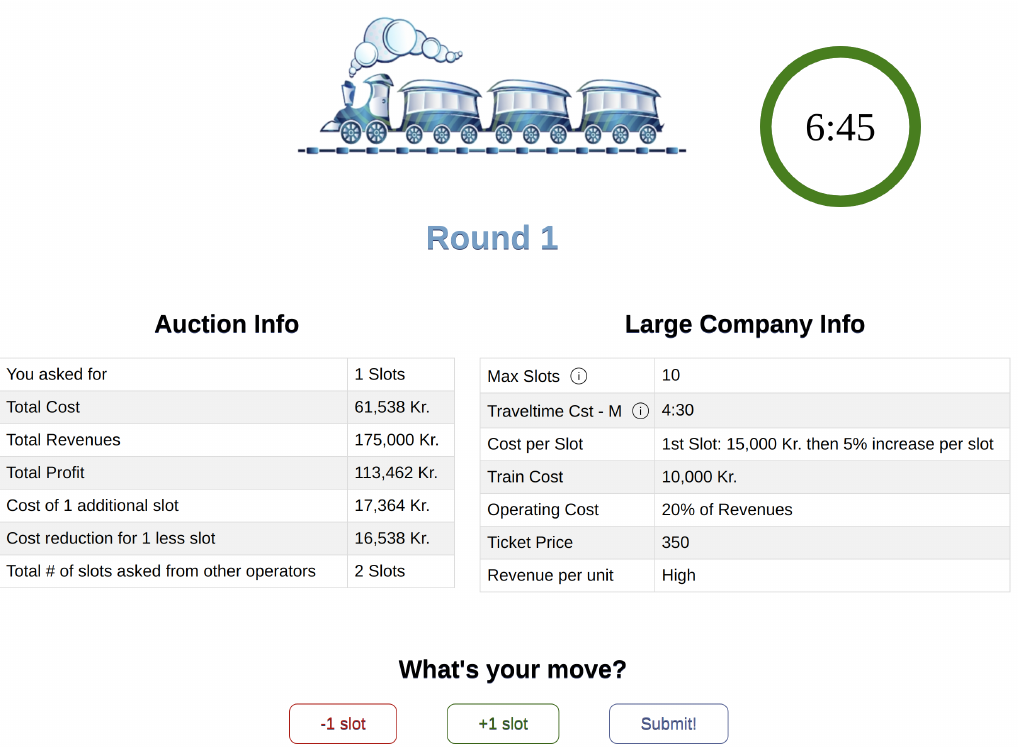}
\caption{The game interface.} \label{fig:gameui}
\end{figure}

The user interface is composed of four main elements, as shown in Figure \ref{fig:gameui}. First, a timer is displayed prominently, indicating the remaining time available for submitting a decision in the current round. Each round is subject to a fixed time limit, and players are required to submit their slot request before the timer expires. Failure to do so results in forfeiting the round, yielding zero slots, revenues, and profits for that round.

Second, the interface presents auction-related information relevant to the operator. This includes the number of slots currently requested, which is the only value that can be modified by the player, as well as calculated values such as total cost, revenues, profit, and the change in cost associated with requesting one additional or one fewer slot. The interface also displays the total number of slots requested by competing operators in the previous round, providing players with feedback on the competitive environment.

Third, static company information is shown, including constraints on the maximum number of slots that can be requested and explanatory information on how financial outcomes are calculated. This information remains constant throughout a game session and is intended to support informed decision-making.

Finally, the interface includes control elements that allow players to increment or decrement their slot request and to submit their final decision for the round. Adjusting the slot request dynamically updates the displayed financial indicators, enabling players to explore the immediate implications of alternative decisions before submitting their choice.

This web-based implementation allows the auction model to be experienced as a real-time, interactive system in which individual decisions directly and immediately affect the outcomes of all participants. It also enables systematic collection of gameplay data, which can be used for exploratory analysis and to inform future refinement of both the auction mechanism and the game design.

While the web-based implementation enables automated gameplay, future iterations may benefit from incorporating facilitation design elements. Research on gaming simulations emphasizes that facilitators play a crucial role in managing learning conditions and supporting participants' sense-making processes \cite{de2025advancing}. In the context of railway capacity auctions, facilitation could support participants in connecting gameplay experiences to real-world policy implications and in reflecting on strategic decision-making patterns.

\section{Exploratory Gameplay Observations}\label{sec:observations}

Following the implementation of the auction game, two structured gameplay sessions were conducted. The aim was to examine how railway operators respond to congestion-based pricing and corrective incentives in a dynamic, repeated-interaction environment. In these game sessions, operators of different sizes repeatedly request railway slots under a pricing mechanism that increases costs with aggregate demand. A reward-penalty structure is incorporated to discourage excessive capacity requests by large operators and to support participation by smaller operators. The game design, as described in Section 3, emphasizes behavioural realism by introducing time constraints on decision-making and by providing participants with real-time feedback on costs, revenues, and marginal pricing effects. The primary objective of these sessions is not to identify equilibrium outcomes, but to observe strategic patterns, dominance-related behaviour, and the extent to which pricing incentives influence slot-request decisions under competitive pressure.

We emphasise that these sessions are exploratory and observational rather than experimental: the same rule set and pricing parameters were held fixed across sessions to ensure consistency, but no conditions were manipulated and no hypotheses were tested. References to ``structured'' sessions denote this standardisation of rules, not a controlled-experiment design.

\subsection{Participants and Gameplay Setup}\label{subsec:setup}

In these game sessions, three operators of different sizes – referred to as a small, a midsize, and a large company – participated in repeated rounds of slot allocation. The operators were played by railway experts from the Swedish infrastructure manager, Trafikverket. Each game session consisted of multiple rounds, and the same auction rules and pricing mechanisms were applied throughout.

The participants had professional experience in railway capacity planning and were familiar with the institutional context of railway operations in deregulated markets. Prior to gameplay, participants were briefed on the auction rules, the congestion pricing formula, and the reward-penalty mechanism. They were informed that slot acquisition costs would increase with aggregate demand and that the operator requesting the highest number of slots would incur a 20\% penalty, while the operator requesting the fewest slots would receive a 20\% discount on slot acquisition costs.
Each game session was conducted using the web-based implementation described in Section \ref{subsec:implementation}. Participants accessed the game interface remotely and interacted in real time. Each round was subject to a fixed maximum time limit of 7 minutes, within which participants were required to submit their slot request. The game interface displayed the number of slots requested by competitors in the previous round, as well as calculated values for total cost, revenues, profit, and marginal cost changes associated with requesting one more or one fewer slot.

\subsection{Data Collection and Analysis}\label{subsec:datacollection}

Throughout the gameplay sessions, the web-based system automatically recorded all participant decisions and associated financial outcomes for each round. The dataset collected includes, for each operator in each round:

\begin{itemize}
    \item The number of slots requested
    \item Total slot acquisition cost
    \item Operational costs
    \item Rolling stock rental costs (if applicable)
    \item Total revenues
    \item Profit or loss for the round
    \item Whether the reward or penalty mechanism was triggered
    \item Whether the participant forfeited the round due to time expiry
\end{itemize}

The collected data were analyzed to identify patterns in operator behavior across rounds, including:

\begin{enumerate}
    \item Responsiveness of slot requests to aggregate demand and congestion pricing
    \item Impact of the reward-penalty mechanism on bidding strategies
    \item Persistence of aggressive or conservative bidding patterns
    \item Influence of time constraints on decision quality
\end{enumerate}

The analysis focused on descriptive patterns rather than statistical hypothesis testing, as the primary objective was to generate exploratory insights that could inform future experimental design and analytical validation of the auction mechanism.

To support interpretation, the qualitative descriptors used in the following section were operationalised against quantities recorded by the system. A slot request was classified as aggressive when an operator requested a number of slots at or near the upper end of its feasible range, and in particular when it maintained or increased its request despite incurring the high-demand penalty or after a low profit, or even loss, in the preceding round. A request was classified as conservative when an operator requested at or near the lower end of its range, frequently positioning itself to receive the low-demand discount. Decision quality was assessed relative to a myopic best-response benchmark: given the competitors' total request observed in the previous round, the profit-maximising slot count is directly inferable from the marginal-cost indicators shown in the interface (the cost of one additional and one fewer slot). A decision was treated as higher-quality when it moved the operator toward this benchmark and lower-quality when it moved away from it or when the round was forfeited through time expiry. Patterns were identified by the authors through round-by-round inspection of the recorded logs, comparing each operator's choices against (i) its own previous-round request, (ii) the reward/penalty thresholds, and (iii) the best-response benchmark. The analysis was descriptive rather than inferential, consistent with the exploratory aim of the study.

\subsection{Findings and Observations}\label{subsec:findings}

As a result of the game sessions, several patterns were identified and observations were made. A first observation concerns the sensitivity of slot prices and costs to aggregate demand. Across gameplay sessions, increases in the total number of requested slots led to noticeable increases in slot acquisition costs for all operators, confirming that the congestion-based pricing mechanism functioned as intended. Players were able to observe these effects in real time through the user interface, which displayed the marginal cost of requesting an additional slot as well as the cost reduction associated with requesting fewer slots.

The reward–penalty mechanism introduced in the second variation of the auction model was also active during gameplay. Operators requesting the highest number of slots in a round incurred higher slot acquisition costs due to the penalty applied, while operators requesting the fewest slots benefited from reduced slot costs. Smaller operators frequently adopted conservative slot-request strategies and thus benefited from the discount, indicating that the corrective incentives were being triggered in practice. At the same time, larger operators often continued to request relatively high numbers of slots despite the penalty, suggesting that the corrective mechanism did not fully deter aggressive capacity requests.

Gameplay data further indicate that operators did not always adjust their slot requests in a strictly profit-maximizing manner from one round to the next. In several cases, operators maintained or increased their requested number of slots even after experiencing low or negative profits in the previous round. This behaviour suggests that strategic considerations, such as maintaining market presence or influencing competitors’ costs, may have played a role in decision-making alongside short-term profitability.

Finally, the presence of time constraints affected gameplay outcomes. Each round was subject to a fixed time limit for submitting decisions, and failure to submit within this limit resulted in forfeiting the round. Such procedural losses occurred during gameplay and led to zero economic outcomes for the affected operator in that round. These cases are best interpreted as decision failures rather than strategic choices and highlight the role of time pressure and cognitive constraints in interactive capacity allocation settings.

Following each session, an informal debrief was conducted in which participants were asked to explain the reasoning behind their decisions. A consistent theme emerged: participants reported that their bidding did not reflect their personal disposition but rather what they judged to be appropriate for the operator they were representing, suggesting that decisions were driven by the assumed role rather than individual temperament. This role-oriented framing is consistent with the intended use of the game as a vehicle for examining operator behaviour rather than personality. The most specific account was offered by the participant representing the large operator, who stated that he was willing to accept a small loss in order to preserve market-leader status and to prevent competitors from securing the more desirable, or ``good'', slots. This explanation aligns closely with the behavioural pattern observed in the gameplay data, namely the persistence of high slot requests by the large operator despite the penalty and occasional negative profits, and lends qualitative support to the interpretation that strategic considerations—such as maintaining market presence and raising rivals' costs—operated alongside short-term profit maximisation.

Overall, these observations provide initial empirical grounding for the proposed auction model and illustrate how congestion pricing, corrective incentives, and time-constrained decision-making jointly shape operator behaviour. The results are exploratory in nature and are intended to inform future analytical validation and more extensive experimental evaluation.

\section{Conclusion \& Future Work}\label{sec:conclusion}

This paper presented an auction-based model for pricing and allocating railway slots, along with a multiplayer game designed to support its exploration and evaluation. The proposed approach combines congestion-based pricing with corrective incentives aimed at discouraging monopolistic behaviour and supporting participation by smaller operators. A prototype implementation was first developed in a spreadsheet environment, followed by a web-based version that enables real-time interaction, time-constrained decision-making, and systematic data collection.

In addition to describing the design and implementation of the auction game, the paper reported initial observations from structured gameplay sessions involving operators of different sizes. These observations indicate that the congestion-based pricing mechanism responds to aggregate demand according to the design and that the reward–penalty scheme is actively triggered during gameplay. At the same time, the persistence of aggressive slot-request strategies by larger operators suggests that corrective pricing alone may not be sufficient to fully neutralize strategic dominance. The observed behaviour further highlights the role of strategic reasoning and procedural constraints, such as time pressure, in interactive capacity allocation settings.

The results presented in this paper are exploratory in nature and are not intended as a comprehensive validation of the proposed auction mechanism. Rather, they serve as an initial empirical grounding of the design and as input for further refinement of both the auction model and the game environment. Future work will focus on analytical validation of the auction mechanism \cite{roungas2017framework,roungas2018framework}, extended experimental studies with domain experts, and more systematic analysis of gameplay data. An informal post-session debrief provided initial qualitative insight into participants' reasoning; future work would benefit from more systematic semi-structured interviews to triangulate participants' stated rationales with the behavioural data. Given that games constitute socio-technical systems where meaning emerges through participant interaction \cite{meijer2025positioning}, future studies should also consider facilitation approaches \cite{de2025advancing} that support participants in transferring insights from gameplay to real-world railway capacity allocation contexts. Ultimately, the goal is to assess the effectiveness of the proposed approach in promoting efficient, competitive, and socially desirable outcomes in railway capacity allocation.

\bibliographystyle{splncs04}
\bibliography{biblio}

\end{document}